\documentclass[12pt]{revtex4}
\usepackage{graphicx}
\usepackage{amsfonts,amssymb,eucal,amsmath}

\numberwithin{theorem}{section}
\numberwithin{equation}{section}
\begin{document}

\title{Multiple-relaxation-time Finsler-Lagrange dynamics
    in a compressed Langmuir monolayer}

\author{V. Balan}
\email{vladimir.balan@upb.ro} \affiliation{ University Politehnica
of Bucharest, Faculty of Applied Sciences, Department of
Mathematics-Informatics, 313 Splaiul Independentei, 060042 Bucharest, ROMANIA.}%

\author{H. V. Grushevskaya}
%
\email{grushevskaja@bsu.by} \affiliation{ Physics Department,
Belarusian State University,
4 Nezavisimosti ave., 220030 Minsk,  BELARUS}%

\author{N. G. Krylova}
%
\email{nina-kr@tut.by} \affiliation{ Physics Department,
Belarusian State University,
4 Nezavisimosti ave., 220030 Minsk,  BELARUS}%

\author{M. Neagu}
\email{mircea.neagu@unitbv.ro}
\affiliation{ Department of Mathematics and Informatics, University Transilvania of
Bra\c{s}ov, 50 Iuliu Maniu Blvd., 500091 Bra\c{s}ov, ROMANIA}%

\date{}

\begin{abstract}
In this paper an information geometric approach has been proposed
to describe
    the two-dimensional (2d) phase  transition of the first order in a monomolecular
    layer (monolayer) of amphiphilic molecules deposited on air/water interface.
The structurization of the monolayer was simulated as an entropy
evolution of a
    statistical set of microscopic states with a large number of relaxation times.
The electrocapillary forces are considered as information
constraints on the
    statistical manifold.
The solution curves of  Euler-Lagrange equations and the Jacobi
field equations
    point out contracting pencils of geodesic trajectories on the statistical
    manifold, which may change into spreading ones, and converse.
It was shown that the information geometrodynamics of the
first-order phase
    transition in the Langmuir monolayer finds an appropriate realization within
    the Finsler-Lagrange framework.
\end{abstract}

\maketitle



\section{Introduction}
The relaxation processes, which depend on their flowing-speed $V$ are
    experimentally observed during phase transitions from an isotropic phase to anisotropic
    2d-phases or to a 3d-phase (collapse state) of compressed monolayers fabricated
    from amphiphilic molecules of surface-active substances.
We call {\em isotropic phase} a liquid-expanded (LE) state.
The tilted liquid condensed (LC) states $L_2$, $L^{\prime}_2$,
$Ov$, and $S$ are related
    to the anisotropic 2d-phases with slant hydrophobic "tails" \ ({\em hydrophobic moiety})
    of the amphiphilic  molecules.
The untilted solid-like anisotropic 2d-phases $L^{\prime\prime}_2$
and $CS$ are called
    {\em solid condensed (SC) states} \cite{1,Tadjer}.
The compressed monolayers from hydrated surface-active substances
extracted from biological
    membranes crystallize at low compression rates $V$ \cite{Postle,Eeman} and collapse
    during quasi-static compression.
Such Langmuir monolayers become metastable and do not collapse at
much higher surface
    pressures when compressed faster than a threshold rate, while the dependence of
    monolayer properties on the compression rate is not determined by the composition
    of the monolayer and, respectively, by different miscibilities of substances \cite{Smith}.
This metastable state is a state of type of supercooled liquid in
the first-order phase
    transition \cite{Landau-statistic}.
The properties of metastable state depend on the action time, similarly to the glass or
    ferroelectric cases. A system can stay arbitrarily long in the supercooled liquid state,
    if the seed crystallizing centers are absent.
Such centers are imperfections, or impurity particles, or new phase elements of the
    above-critical size.
A {\em supercooled-liquid heterogeneous dynamics} is a dynamics of a system consisting from
    individual relaxing units which have site-specific relaxation times and, hence, are
    characterized by different coexisting timescales \cite{Richert}.
Similar multiple-relaxation-times heterogeneous dynamics depending
on compression speed
    holds in a Langmuir film of gold nanoparticles with diameter less than 7~nm, because
    there exist effects of the speed on the compression process, such as a non-linear
    dependence of the loss absolute value (the product between the viscosity of the
    Langmuir monolayer and the compression rate $V$) and the loss factor on $V$ \cite{4}.\par
The  Langmuir monolayer structure is formed from hydrate complexes
of surfactant, because
    a water subphase abundance is expelled from compressed monolayer to outside.
This process is revealed as the occurrence of water drops on the side of hydrophobic tails.
These drops were observed as hill-like structures when the tilted LC 2d-phases or the
    untilted SC \cite{hill-structures} are formed.
The tails in the tilted phases are not arranged vertically, and because of that, unlike the case of
    untilted phases, the hydrophobic interactions prevent the return of expelled water
    \cite{hill-structures}.
This ejection process depends on the speed  $V$ of  monolayer compression.
The drops of ejected subphase particles were also observed as subsided bubbles in
    ultrathin Langmuir-Blodgett (LB) films \cite{LB-magnit_SPIE}.
If the LB-film is thick (9 and more monolayers), the water from
the bubbles is retained via
    interlayer interactions and the bubbles do not fall \cite{LB-magnit_SPIE}.
The dependence of hydrate-complex stability on the compression speed $V$  for a monolayer
    formed from polymers such as Isotactic and Syndiotactic Poly(methyl methacrylate) at a dilute
    state can explain the dependence of the blending ratio on $V$, which was observed in
    \cite{Stereocomplexation}.
These substances - being miscible at the rates $V\sim 0.001$~mm/s - are not miscible if $V=0$.
The width of the plateau of isotherms for stearic acid also depends on the value of
    the compression rate $V$ \cite{3,myJ.Phys.CS2015}.
The most commonly used recent models of phase transitions in layered systems are the
    {\em statistical lattice models} \cite{15,Chamberlin1993,Chamberlin1999}.
Such an approach is based upon mean-field theory \cite{Landau-phys-kin}.
The phase nuclei (phase elements, relaxing  units) may appear/disappear with a given
    probability at each lattice location, when the monolayers are structuring.
These phase elements have their own lifetimes (relaxation times) \cite{Skripov}.
A theory of 2d-phase transitions was developed for a model of interacting tails of
    amphiphilic  molecules, where electrostatic repulsion of polar "heads" \
    (hydrophilic  moiety) of these molecules is considered as a small perturbation     \cite{Kaganer}.
In this theory, the free energy is varied under the assumption of homogeneity of
    the phase elements, and one completely neglects the existence of the metastable
    transient state.
Besides this, the head-enveloping water molecules, whose number depends on the geometry
    at certain area per molecule, play a crucial role for the overall electric properties
    of the monolayer \cite{Tadjer}.
The free energy of a system with such a feature - like the presence of a broad spectrum
    of relaxation times of heterogeneous phase elements, that is characteristic for the
    metastable state - is also varied in the Kolmogorov-Mehl-Johnson-Avrami approach to
    the first-order phase transition \cite{Avrami,Johnson,Kolmogorov}.
This approach, however, does not give any information about the transient phase-element
    kinetics \cite{Shur1996}.
Since the metastable state is an unstable one with a negative compressibility, it can be
    represented as a collection of a large number of elementary unstable
    independently relaxing units, for example, bistable elements in a Preisach  model \cite{Bartic,Cima1,Cima2}.
The use of the Preisach  model, which was initially proposed to describe the ferromagnetic case,
    allows to take into account an entropy contribution, which is due to the random polarization
    of the environment, as  "coercive forces"\ and, respectively,  gives a statistical distribution of
relaxation times.
However, it was shown in \cite{Shur2007}, that an appropriate description of the metastable state
    is impossible without variation of the entropy contribution, particulary,
    without variation of the coercive forces     \ in the Preisach  model.\par
The evolution of a discrete set of molecule orientation parameters, which describe the crystal
    space rearrangement, is examined in the hydro-dynamic (continuous) limit of a kinetic theory
    of the  second-order phase transition from the tilted allotropic crystalline  form to the untilted
    one \cite{Kaganer}.
Then, the powerful field theory and the renormalization group methods are used to describe
    the second-order phase transitions.
The continuous parameterizations with space-scaling  are not applicable for  first-order
    phase transitions, because the matestable state has to be scaled in the time domain, but not
    in the space one \cite{Sh.-Xi.Qu2000}.\par
The first-order phase transition in Langmuir monolayers - as  process with relaxation-time dispersion -
is characterized by a statistical distribution, that is fluctuated between
    the statistical distributions $p_L$ and $p_S$ for amphiphilic molecules in the expanded
    liquid and in the crystalline phases, respectively.
We shall further use the {\em information geometrodynamics}
\cite{6,9,10,11,Amari2000}
    and shall variate the entropy in order to describe the 2d-phase transition of the first order.
The information geometric techniques will be utilized to analyze
the stability of the most
    probable trajectories on the statistical manifold, which consists of the set of probability
    distributions $\{p_i\}$ for the $i$-th microscopic states (microstates) under certain
    thermodynamic conditions regarded as information constrains.
The statistical manifolds can be non-trivial tangent bundles \cite{Shen2006}.
The dependence of the structurization process in the Langmuir monolayers on the compression rate
    means that an addition of compression-velocity vector to a tangent space of the monolayer
    may change the sub-spaces formed by sets of tangent vectors at points of the arbitrary
    trajectory in the coordinate space of the monolayer.
Then, there exists a non-trivial slice made of tangent unit vectors, which form the so-called
    indicatrix surface over the (base) coordinate space.
Accordingly, a heterogeneous dynamics of the structurization occurs in the tangent bundle.
Therefore, in our paper we will use Finsler geometry structures
\cite{18,14,DifGeomDynSys2015}
    of tangent bundles to describe the Langmuir monolayer structurization as a process
    that is characterized by a dynamic scaling law for the relaxation times and by the
    presence of a distribution of timescales.\par
The goal of this article is to construct a statistical manifold of
compressed monolayers,
    which are deposited on the air/water interface, and  the  geometrodynamics of this manifold,
    and to study  the   structurizing  monolayers in transient state of the first-order phase transition.
%
%
\section{The statistical manifold for first-order $2d$-phase transitions}
We assume that the phase nuclei (phase elements) in Langmuir
monolayers are amphiphilic molecules
    in two states: the hydrated complexes and the molecules leaving the complexes (in a free state).
A microstate introduced as the phase element of the 2d-membrane is
a continuous analog of a phase
    nucleus from the ordinary theory of the first-order phase transitions.
An evolution of macroscopic states (macrostates) in the
first-order 2d-phase transition is a
    decreasing (increasing) surface tension due to the decay (production) of the phase elements
    in a compressed Langmuir monolayer.\par
In order to obtain the differential equations which describe the
dynamics of the system on the
    statistical manifold, we shall use the maximum entropy principle \cite{6} and the maximum
    entropy production principle \cite{9,10} in the form proposed in \cite{11}.
Let us consider the distribution
$p(\vec{r}_1,\;\vec{r}_2,\ldots,\;\vec{r}_{N};\,t)$,
    $N\to \infty$ for $N$ microstates ($N$ phase-elements) with the coordinates
    $\{\vec{r}_1,\;\vec{r}_2,\,\ldots,\;\vec{r}_{N}\}$ and the eigenfrequencies
    $\{\omega_1,\;\omega_2,\,\ldots,\;\omega_{N}\}$, $N\to \infty$.
These $N$  microstates arise at the moments $ t_i,\ i=1,\ldots, N$
and decay during the periods
    $\Delta t_i,\ i=1,\ldots, N$ with relaxation times $\{\tau_1,\;\tau_2,\,\ldots,\;\tau_{N}\}$.
Therefore, one has to replace the probability distribution
    $p(\vec{r}_1,\;\vec{r}_2,\ldots,\;\vec{r}_{N};\,t)$  with the following expression:
    $$p(\vec{r}_1,\, t_1+\Delta t_1;\;\vec{r}_2,\,t_2+\Delta t_2;\;\ldots,\;
    \vec{r}_{N},\, t_{N}+\Delta t_{N}),\;\;N\to \infty.$$
The evolution of entropy $S$ reads
    \begin{equation}\begin{split}S(t_i+\Delta t_i)-S(t_i)\\[1mm]
    =-\int p(\vec{r}_1,\, t_1+\Delta t_1;\;\vec{r}_2,\, t_2
        +\Delta t_2;\;\ldots,\;\vec{r}_{N},\, t_N+\Delta t_{N})\\
    \times\ln\frac{p(\vec{r}_1,\, t_1+\Delta t_1;\;\vec{r}_2,\, t_2
        +\Delta t_2;\;\ldots,\;\vec{r}_{N},\, t_N+\Delta t_{N})}
    {p(\vec{r}_1,\,t+\Delta t_1;\; \ldots;\;\vec{r}_{i},\, t_i;\;\ldots,\;
        \vec{r}_{N},\,t_N+\Delta t_{N})}dr_1\ldots dr_{N}.\end{split}
    \label{entropy}\end{equation}
Here the entropy $S(t_i+\Delta t_{i} )$ and $S(t_i )$ at the moments $t_i+\Delta t_{i}$
    and $t_i$, respectively, are defined as
\begin{eqnarray}
&& S(t_i+\Delta t_{i} )=S\left( p(\vec{r}_1,\, t_1+\Delta t_1;\ldots;
    \vec{r}_{i},  t_i+\Delta t_{i};\;\ldots,\;\vec{r}_{N}, t_N+\Delta t_{N})\right),\\[1mm]
&&    S(t_i)\nonumber \\
&&    = S\left( p(\vec{r}_1,\, t_1+\Delta t_1;  \ldots; \vec{r}_{i-1}, t_{i-1}+\Delta t_{i-1};
        \vec{r}_{i},  t_i; \vec{r}_{i+1},  t_{i+1}+\Delta t_{i+1}; \ldots,
        \vec{r}_{N}, t_N+\Delta t_{N})\right) .\quad \nonumber\\
\end{eqnarray}
An information about the macrostate imposes constraints on the distribution of the microstates.
This additional information is an increment $\Delta U_s^i$ of the 2d-membrane free energy
    $ U_s^i$ for the $i$-th phase element of the membrane, due to the change of the electric
    potential difference on the interface between air and water subphases in electrocapillarity
    phenomena \cite{Lippman,Frumkin,LandauVol8,Gibbs}.
All the relaxation processes for all the $N$ phase elements give a contribution $\Delta U_s$ into $ U_s$.
Therefore, $\Delta U_s$ is given by the following expression:
\begin{equation} \label{GrindEQ__2_}
    \Delta U_s =\sum_{i=1}^{N}\Delta \sigma_{i}^{electrocap} A_{t_{i} } \, \Delta
    t_{i}.
\end{equation}
Here $\Delta \sigma_{i}^{electrocap}$ is a surface-tension
increment owing to an
    action of the electrocapillary forces during a small time interval $\Delta t_{i}$
    for  $i-$th phase element, $A_{t_{i} } =\frac{\Delta A}{\Delta t_{i} }<0$ is the
    rate of change of area $A$ per molecule, for the $i-$th phase element.
A constraint stipulated by the adding the information
\eqref{GrindEQ__2_} regarding the
    macrostate of the system has the following form:
\begin{eqnarray} \label{GrindEQ0_3_}
&&\Delta s\int (H+i\Gamma)p(\vec{r}_1,\,t_1+\Delta t_1;\;
    \ldots;\;\vec{r}_{i},\,t_i  +\Delta t_{i};\;\ldots;\;\vec{r}_{N}, \,
    t_N+\Delta t_{N})dr_1\ldots dr_{N} \nonumber\\
&&    =-\sum_{i=1}^{N}\Delta \sigma_{i}^{electrocap}A_{t_{i} } \, \Delta t_{i} \,, \quad \end{eqnarray}
where $H$ is the Hamiltonian, $\Gamma $ is the damping of the system, and $\Delta s$ is the
    increment of the evolution parameter $s$.
By applying the Laplace transform in the left hand side of the expression \eqref{GrindEQ0_3_},
    we can rewrite it as
\begin{eqnarray}\label{GrindEQ0_4_}
    &&\Delta s\int (H+i\Gamma)p(\vec{r}_1,\,t+\Delta t_1;\; \ldots;\;\vec{r}_{i},\,
        t+\Delta t_{i};\;\ldots;\;\vec{r}_{N},\, t+\Delta t_{N})dr_1\ldots dr_{N}\nonumber\\
    &&\quad=\Delta s\sum_{i=1}^{N}\int({\rm\omega }_{i}+i{\rm\tau }_{i}^{-1})  p(\vec{r}_1
        ({\rm\omega }_1 ),\;\ldots\;\vec{r}_{i} ({\rm\omega }_{i} ),\; \ldots,\;\vec{r}_{N}
        ({\rm\omega }_{N} ))dr_1 ({\rm \omega }_1 )\ldots dr_{N} ({\rm\omega }_{N} )\nonumber\\
    &&\quad=-\sum_{i=1}^{N}\Delta \sigma_{i}^{electrocap} A_{t_{i} } \, \Delta t_{i} \, .\end{eqnarray}
Since the relations ${\omega }_{i}{\tau }_{i} \gg 1$, $N\to \infty$ hold for the
    phase transition of the fist order, by using the expression \eqref{GrindEQ0_4_}
    and the normalization condition
\begin{eqnarray}\int p(\vec{r}_1 ({\rm\omega }_1 ),\;\ldots\;\vec{r}_{i}
    ({\rm\omega }_{i} ),\; \ldots,\;\vec{r}_{N} ({\rm\omega }_{N} ))\;
    dr_1 ({\rm\omega }_1 )\ldots dr_{N} ({\rm\omega }_{N})=1,\label{probability_normalization}
\end{eqnarray}
one gets the relaxation times $\tau_{i} $, $i=1,\ldots, N$:
\begin{equation}\label{GrindEQ0_5_}\tau_{i} =-\frac{\Delta t_{i} }{\Delta s} .\end{equation}
By varying \eqref{entropy} jointly with the constraints \eqref{GrindEQ0_4_} and
    \eqref{probability_normalization} multiplied with the Lagrange multipliers $\lambda_{{\rm 1}}$
    and $\lambda_2 $, respectively, we find a non-stationary statistical distribution to which
    our system evolves during the time interval $\Delta t_{i} $:
\begin{eqnarray}\label{GrindEQ0_6_}
&&
p(\vec{r}_1,\, t_1+\Delta t_1;\; \ldots;\;\vec{r}_{i},\,t_i  +\Delta t_{i};\;
      \ldots;\;\vec{r}_{N},\, t_N+\Delta t_{N}) \nonumber \\[1mm]
&&
=p(\vec{r}_1,\, t_1+\Delta t_1;\; \ldots;\;\vec{r}_{i-1},\, t_{i-1}
        +\Delta t_{i-1};\;\vec{r}_{i},\, t_i;\;\vec{r}_{i+2},\,t_{i+2}
        +\Delta t_{i+2};\; \ldots;\;\vec{r}_{N},\,t_N+\Delta t_{N}) \nonumber \\
&&
\times\frac{e^{\Delta s{\kern 1pt} \lambda_2 ({\rm\omega}_{i}
        +i{\rm\tau }_{i}^{-1})} }{Z_{i} },
\end{eqnarray}
where $Z_i$ is the statistical sum
\begin{eqnarray}\label{i-th_StatSum}
&&    Z_{i} =\int e^{\Delta s{\kern 1pt} \lambda_2 ({\rm\omega }_{i}+i{\rm\tau }_{i}^{-1})} \nonumber\\
&&    \times p(\vec{r}_1,\, t_1+\Delta t_1;\; \ldots;\;\vec{r}_{i-1},\,t_{i-1}
        +\Delta t_{i-1};\;\vec{r}_{i},\, t_i;\;\vec{r}_{i+2},\,t_{i+2}+\Delta t_{i+2};\;
        \ldots;\;\vec{r}_{N},\, t_N+\Delta t_{N})dr_{i}. \nonumber
        \\
\end{eqnarray}
Hence, one can determine the statistical manifold $\cal{M}$ of the
examined system with the given
    set of relaxation times \eqref{GrindEQ0_5_} regarded as a set of probabilities \eqref{GrindEQ0_6_}:
\begin{eqnarray}{\cal M}=\{p(\vec{r}_1  (\tilde t_1), \ldots,\vec{r}_{i}
    ( \tilde t_{i-1}), \vec{r}_{i} ( t_i),\vec{r}_{i} ( \tilde t_{i+1}),
    \ldots, \vec{r}_{N}(\tilde t_N))e^{\Delta s \lambda_2 ({\rm\omega}_{i}
    +i{\rm\tau }_{i}^{-1})}/Z_i\},
    \label{stat-manifold_0}\end{eqnarray}
where  $\tilde t_i=t_i+\Delta t_{i},$ $i=1,\ldots, N$.\par
Let us suppose that $\sum_{i=1}^{N}\Delta t_{i}$ be equal to a
phase-transition time $T_{{\rm pht}} $:
\begin{eqnarray}\label{transition_time}\sum_{i=1}^{N}\Delta t_{i} =T_{{\rm pht}}.\end{eqnarray}
According to the expressions \eqref{GrindEQ0_5_} and \eqref{transition_time}, the value
    of the parameter $s$ is defined by the phase-transition time $T_{\mbox{pht}}$ as:
    $\Delta s=-\frac{T_{{\rm pht}}}{\sum_{i=1}^{N}\tau_{i} } $.
Then, assuming that $\lambda_2=\sum_{i=1}^{N}\tau_{i}$ and
performing subsequent
    iterations of the type described in \eqref{GrindEQ0_6_}, we eventually find
    the distribution function during the phase transition, e.g., from the liquid
    expanded state with the distribution function $p_{L}$ to the crystalline condensed
    one with the distribution function $p_{CS}$:
\begin{eqnarray}\label{distribution_function}p_{CS} (\vec{r}_1,\;\vec{r}_2,
    \ldots,\;\vec{r}_{N};\, t+T_{{\rm pht}}) =p_{L} (\vec{r}_1,\,\;\vec{r}_2,\,
    \ldots,\;\vec{r}_{N};\,t)\;\frac{e^{-(H+i\Gamma)T_{{\rm pht}} } }{Z},\end{eqnarray}
where the statistical sum \textit{Z} is defined by the expression:
\[Z=\int \prod_{i=1}^{N} dr_{i} p_{ L} (\vec{r}_1,\,\;\vec{r}_2,\,
    \ldots,\;\vec{r}_{N};\, t)e^{-T_{{\rm pht}} (H+i\Gamma)} .\]
%
%
\section {Information geometrodynamics of the statistical manifold}
By using \eqref{GrindEQ0_6_}, we will construct the
geometrodynamic approach with
    the electrocapillary forces regarded as information constraints to describe the
    first-order 2d-phase transition in the compressed Langmuir monolayer.
The electrocapillary part of free energy \eqref{GrindEQ__2_} is
determined by the
    following Lagrange function \cite{16}:
\begin{equation}L(t,r,\dot{r},\dot{\varphi})=\frac{m}{2}\dot{r}^2+\frac{mr^2}{2}\dot{\varphi}^2
    \underset{\overset{\shortparallel }{U_{s}(t,r)}}{\underbrace{-\tilde{p}r^5|V|e^{\frac{2|V|t}{r}}\cdot
    \dot{r}^{-1}+U(t,r)}},\label{2D-Lagrangian}\end{equation}
where $m$ is the mass of the particle, $V$ is the compression
speed,
    $\tilde{p}=\frac{\pi ^2q^2}{\varepsilon \varepsilon_0}\frac{\rho_0^2}{R_0^2}$ is a constant
    which depends on the molecular parameters of the monolayer;
    $U_{s}(t,r)$ is the potential of the electrocapillary forces:
\begin{eqnarray}
U(t,r) &=&\tilde{p}\left\{ \left[
-\frac{4}{3}r^5+\frac{16}{15}(|V|t) r^4
    +\frac{1}{30}( |V|t)^2r^3+\frac{1}{45}(|V|t) ^3r^2\label{u} \right.\right.\\
    &&
    \left. \left. +\frac{1}{45}( |V|t)^4r+\frac{2}{45}( |V|t) ^5\right]e^{\frac{2|V|t}{r}}
    -\frac{4}{45}\frac{( |V|t)^{6}}{r}\,\mbox{Ei}\left( \frac{2|V|t}{r}\right)\right\},\notag
\end{eqnarray}
and $\mbox{Ei}=-\int_{-z}^{\infty}\frac{e^{-t}}{t}dt$ is a special
function.\par
Then, one can reconsider the statistical manifold $\cal{M}$
\eqref{stat-manifold_0}, as
\begin{eqnarray}{\cal M}=\{p(\vec{r}_1, t_1+\Delta t_1; \vec{r}_2, t_2
    +\Delta t_2; \ldots; \vec{r}_{N}, t_N+\Delta t_{N}|\left<{\rm\omega }_{i}
    {\rm\tau }_{i}\right>\propto L_i \Delta t_i/\Delta s, \tau_i
    =-\Delta t_i /\Delta s)\},\nonumber \\ \label{stat-manifold}\end{eqnarray}
where $L_i\equiv L(t_i,r_i,\dot{r}_i,\dot{\varphi}_i)$.\par
Each microstate is a point on the geodesics of the statistical manifold ${\cal M}$
    \eqref{stat-manifold}, which are parameterized by the macrostates $L_i\tau_i$.
%
%
The time set $\{t_k, \Delta t_m \}$ determines an $N^2$-dimensional macroscopic
    vector $\Theta$ with statistical coordinates $\left\{\theta^{(m)}_k\right\} $,
    where $k = 1, 2, \ldots, N$ label the microstates and $m = 1, 2, \ldots , N$
    enumerate the information constraints.
According to the expressions \eqref{entropy} and
\eqref{stat-manifold}, the variation
    of entropy $\delta S$ is determined by the following expression:
\begin{eqnarray}\delta S =\sum_a\left[\int \ln p(X) {\partial p(X)
    \over\partial\Delta t_a}\delta (\Delta t_a) dX + \int p(X){\partial \ln p(X)
    \over\partial \Delta t_a}\delta (\Delta t_a)dX \right],\label{entropy_variation0}
\end{eqnarray}
where  $X=\{\vec{r}_1  (\tilde t_1), \ldots, \vec{r}_{N}(\tilde
t_N)\}$ is a configuration.\par
By substituting the expression \eqref{distribution_function} into
the equation
    \eqref{entropy_variation} and taking into account \eqref{stat-manifold},
    one obtains the variation of entropy for the Langmuir monolayer in the phase transition:
\begin{eqnarray}\delta S =\sum_a\left[\int \ln p(X) {\partial p(X)
    \over\partial\Delta t_a}\delta (\Delta t_a) dX\right]+\sum_a \tau_a\delta s
    \int p(X)(H+i\Gamma)   dX . \label{entropy_variation}
\end{eqnarray}
By introducing generalized momenta $P(X)$ and generalized velocities $\dot X$ on
    ${\cal M}$, by using the relation $P\dot X = T_{pht}(H+i\Gamma)^2$
    one can rewrite \eqref{entropy_variation} in the following  form:
\begin{eqnarray}\delta S =  \sum_a \tau_a\delta s\int p(X)\left[(H+i\Gamma)
    - P(X)\dot X  \right]dX . \label{entropy_variation1}\end{eqnarray}
According to the expressions \eqref{stat-manifold}, \eqref{entropy_variation0}, and
    \eqref{entropy_variation1}, the length $\Delta l$ of the configuration (path, trajectory)
    in the statistical manifold
    $${\cal M}=\{X|\sum_i p_{X_i}(H_i+\Gamma_i)
    \propto \sum_i L_i \Delta t_i/\Delta s, \tau_i =-\Delta t_i /\Delta s)\}$$
    is determined by the following expression:
\begin{equation}\Delta l\propto -\sum_{i=1}^{N} L_i\Delta t_i .\label{GrindEQ0_8_}\end{equation}
The variation condition $\delta S = 0$ gives us a state with maximum entropy,
    and hence the solution curves of the Euler-Lagrange equations with the
    Lagrangian $L_i$ entering in (\ref{GrindEQ0_8_}) is a most probable path on ${\cal M}$.\par
Now we can pass to the continuous-medium limit by replacing the finite decrement
    of the variables in \eqref{GrindEQ0_5_} with the following differentials:
\begin{equation}\dot{\xi }_{s} =\frac{\partial t}{\partial s}
    =\lim\limits_{\Delta s\to 0} \frac{\Delta t}{\Delta s} .\label{GrindEQ__9_}
\end{equation}
At the continuous limit \eqref{GrindEQ__9_}, the expression
\eqref{GrindEQ0_8_} becomes the action
\begin{equation}
    \Delta l \propto -\int { L}\, \dot{{\rm \xi }}_{s} \, ds\label{GrindEQ__10_}
\end{equation}
on the statistical manifold ${\cal M}$, with the metric function
$dl$:
\begin{equation}
    dl\propto -{ L}\, \dot{{\xi }}_{s} \, ds.\label{GrindEQ__11_}
\end{equation}
%
\section{The Finsler-Lagrange dynamics of the Langmuir monolayer: numerical modeling}
Let us assume that all the phase elements decay with the
relaxation time
    $\dot{{\xi }}_{s}=1$ during a small time-interval.
In this case, the information geomerodynamics of the particles
from the compressed
    monolayer during small time-intervals is described by the action \eqref{GrindEQ__11_}
    with $s=t$.
Therefore the Finsler-Lagrange space of the monolayer in polar
coordinates $(r,\;\varphi)$
    can be described by the following non-relativistic action
\begin{equation} dl=mc^2 dt - Ldt\label{(55)},\end{equation}
where $L$ is the Lagrange function \eqref{2D-Lagrangian}.
Now, one can get the {\em 2-dimensional most probable trajectory}
    (mean trajectory) of the particle in the monolayer.
By making zero the variation of the action (\ref{(55)}), one
obtains the following system
    of differential equations for geodesics (the Euler-Lagrange equations):
\begin{equation}{\dfrac{dy_1^{i}}{dt}}+2G_{(1)1}^{(i)}( t,x^{k},y_1^{k})=0,\qquad
    \dfrac{dx^{k}}{dt}=\dot{x}^{k}\equiv y_1^{k},\label{geodesic-equation}\end{equation}
where
\begin{equation}\begin{split}G_{(1)1}^{(1)}=\dfrac{\tilde{p}r^3|V|e^{\frac{2|V|t}{r}}( 5r\dot{r}^{-1}
    -2|V|t\dot{r}^{-1}+|V|r\dot{r}^{-2})
    -\dfrac{1}{2}\dfrac{\partial U}{\partial r}-\dfrac{mr}{2}\dot{\varphi}^2}{m
    -2\tilde{p}r^5|V|e^{\frac{2|V|t}{r}}\cdot \dot{r}^{-3}},\qquad
    G_{(1)1}^{(2)}=\dfrac{\dot{r}}{r}\dot{\varphi}.\end{split}\label{add2to-geodesic-equation}
\end{equation}
Now we shall show that the compression speed $V$ of the monolayer determines the magnitude
    of cohesive force (friction), which acts on the monolayer moving on the subphase surface.
If $V=0$, then the system of differential equations
(\ref{geodesic-equation},
    \ref{add2to-geodesic-equation}) reduces to:
\begin{equation}-r\dot{\varphi}^2+20\tilde{p}r^4/3m+{\ddot{r}}=0,\quad
    2{\dot{\varphi}}{\dot{r}}/r+\ddot{\varphi}=0.\label{V=0geodesic}\end{equation}
Since $V=0$, a solution of the equation \eqref{V=0geodesic} is a most probable trajectory in absence of friction.
In this case, the monolayer is a conservative system.
The shape of geodesic trajectories (\ref{V=0geodesic}) from Fig.~1a shows that the molecules
    of the monolayer in average move around the reference point, along the extended orbits
    revolved to each other, and periodically pass from one orbit to another.
Thus, if $V=0$, the  molecules remain in the hydrated complex, and the system behaves itself
    like a conservative one.\par
According to Fig. 1b, the  monolayer - compressed at small $V$ - may be represented as a system
   with a weak dissipation and, respectively, the trajectories are of limit-cycle type.
A mean trajectory of the particle for large $V$ exhibits inflexion points, as shown in Fig.~1c.
For this reason, such paths may describe metastable states of the system in the first-order phase transition.
\begin{figure}[htbp]
\begin{center} (a)\hspace{10cm}(b) \\
    \includegraphics[width=5.5cm,height=5.5cm,angle=0]{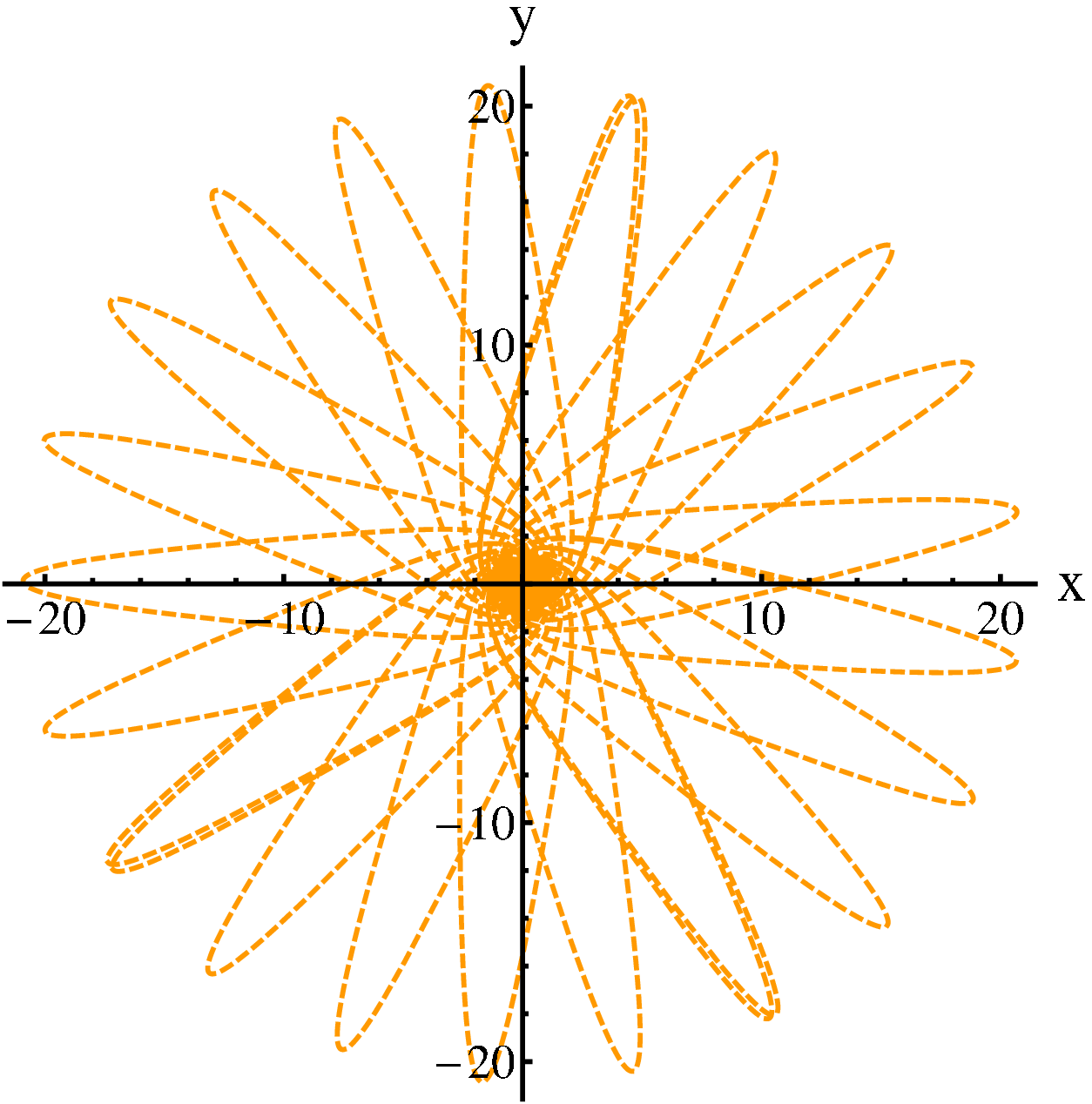}
    \includegraphics[width=8.5cm,height=5.0cm,angle=0]{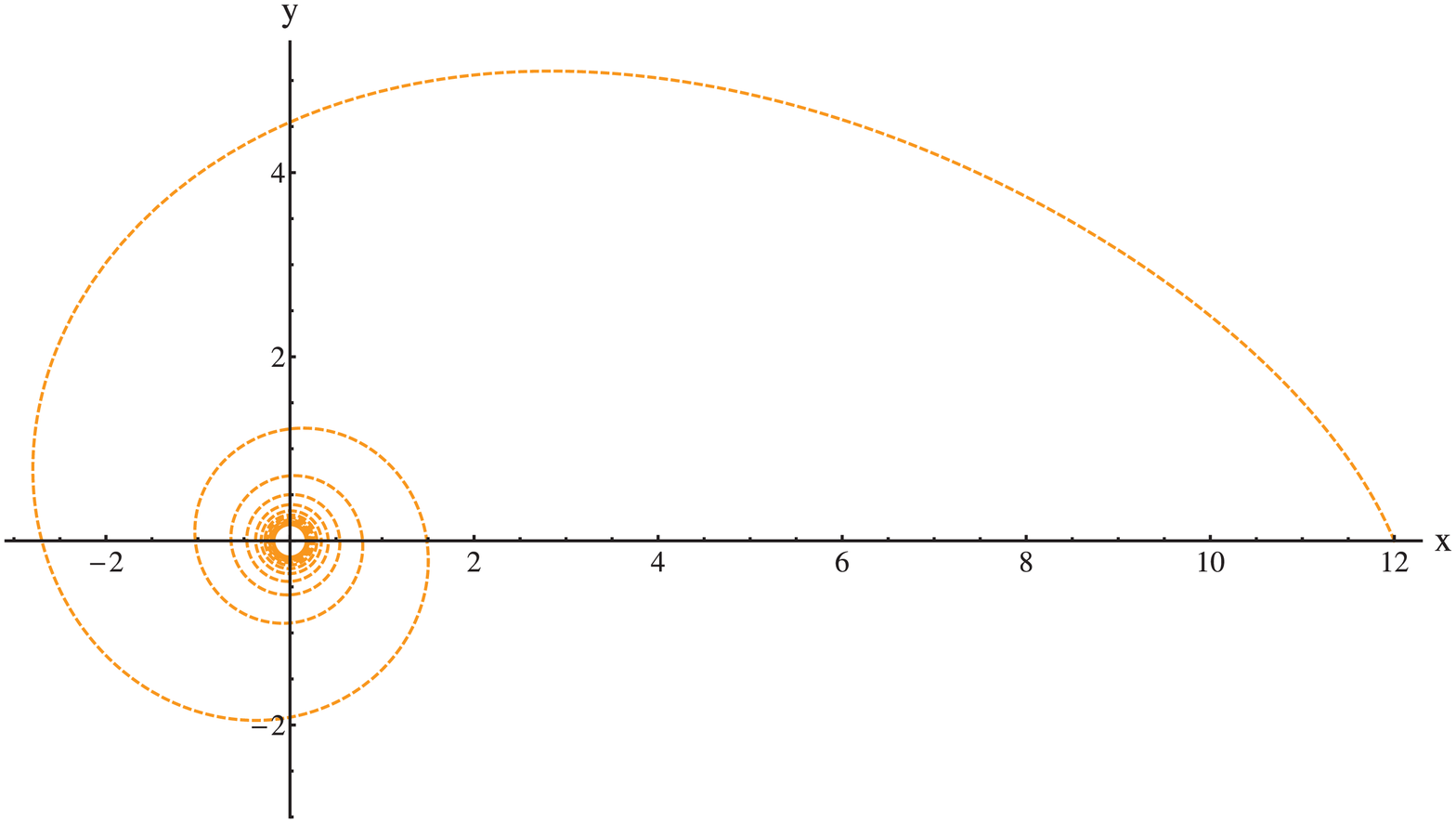}\\
    (c)\\
    \includegraphics[width=10cm,height=4.0cm,angle=0]{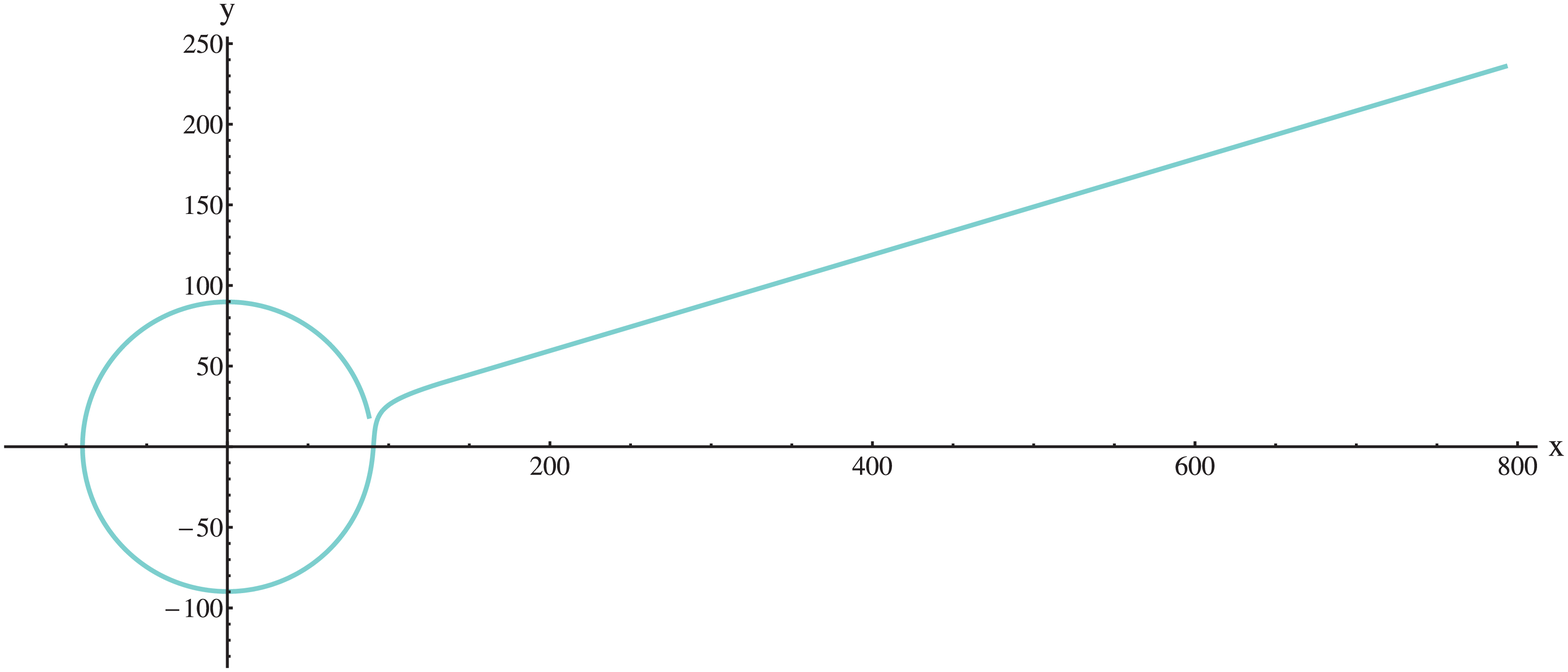}\\
\caption{
    The mean trajectories $\vec r$ of the particle  in the compressed monolayer at $\tilde{p}=10$, $m=1$
    for different compression speeds: (a) $V=0$; (b) $V=1$; (c) $V=500$.}
\end{center}
\end{figure}
We shall further investigate the structure of the set of inflexion points, in the sense of
    Jacobi stability by means of the KCC (Kosambi-Cartan-Chern) invariants theory \cite{18}.
%
%
\section{ Jacobi equations for deviations of geodesics from an instanton-like geodesic}
A system of differential equations for the variation $(\delta r, \delta \varphi)$
    (which describe Jacobi fields) of the geodesics satisfying the system
    \eqref{geodesic-equation} has the following form \cite{16}:
\begin{eqnarray}\nonumber\delta \ddot{r}= \frac{\tilde{p} \dot{r} }{45 r^3
        (2\tilde{p}V r^5 e^{\frac{2Vt}{r} }+m \dot{r}^3)}( 8V^{6} t^{6}\mbox{Ei}\left(\frac{2Vt}{r}\right)
        \dot{r}^2\delta r+e^{\frac{2Vt}{r} } r (45V r^5(5r-2Vt)\delta \dot{r}\\
    -\delta r (4V^5 t^5 +2V^4 t^4 r+182V^3 t^3 r^2 -807V^2 t^2 r^3 +1536Vtr^4
        -1200r^5)\dot{r}^2)),\label{GrindEQ__6_}\\
    \label{GrindEQ__7_}\delta \ddot{\varphi}=0.\end{eqnarray}
In order to study the Jacobi stability, one can assume that the trajectory with
    inflexion points from Fig.~1c is located in the vicinity of an instanton-like geodesic.
A system of differential equations for the  instanton-like geodesic has the following form \cite{16}:
\begin{eqnarray}\dot{r}^3+6 \frac{\tilde{p}V}{m} r^{5} e^{\frac{2Vt}{r}}=0, \ \ \ \
    \dot{\varphi}-C_{0}/r=0,\label{instanton_0}\end{eqnarray}
where $C_0$ is a constant.
To obtain a solution of the system \eqref{instanton_0} for large
enough compression-times,
    one can simplify the system, by taking into account that $r\to(R_{0}-Vt)$ for $t\to\infty$.
Then the system \eqref{instanton_0} takes a following form:
\begin{eqnarray}\dot{r}^3+6 \frac{\tilde{p}V}{m} r^{5} e^{\frac{2Vt}{R_{0}-Vt}}=0 , \ \ \ \
    \dot{\varphi}-C_{0}/r=0,\label{GrindEQ__7a_}\end{eqnarray}
a solution of which is
\begin{eqnarray}
&&r={27m^{\frac{1}{2}}R_{0}V e^{-\frac{Vt}{R_{0}-Vt}}} \nonumber \\
 &&   \left\{\left(6^{\frac{4}{3}}\tilde{p}^{\frac{1}{3}}R_{0}^{\frac{5}{3}}
        +9m^{\frac{1}{3}}V^{\frac{2}{3}}\right)e^{-\frac{2}{3}\frac{Vt}{R_{0}-Vt}}
        -6^{\frac{4}{3}}\tilde{p}^{\frac{1}{3}}R_{0}^{\frac{2}{3}}\left(R_{0}-Vt\right)+4\cdot
        6^{\frac{1}{3}}\tilde{p}^{\frac{1}{3}}R_{0}^{\frac{5}{3}}e^{-\frac{2}{3}\frac{R_{0}}{R_{0}-Vt}}
        \right. \nonumber \\
&&    \times \left.\left(\mbox{Ei}\left(\frac{2}{3}\right)+\mbox{Ei}\left(\frac{2}{3}
        \frac{R_{0}}{R_{0}-Vt}\right)\right)\right\}^{-1}\label{instanton}\end{eqnarray}
Taking into account the expression \eqref{instanton}, a solution of  the system
    (\ref{GrindEQ__6_}, \ref{GrindEQ__7_}) gives a mean trajectory
    $(r(t)+\delta r(t),\varphi (t)+\delta \varphi(t))$ of motion for a particle
    in the monolayer.
A result of the  numerical calculation of the Jacobi fields is presented in Fig.~2 in
    Cartesian coordinates $(x(t),y(t))$.
The solution  shows that the contraction of pencils of geodesic trajectories on the
    statistical manifold interchanges by spreading, and converse.
\begin{figure}[htbh]
\begin{center}
    \includegraphics[scale=.7]{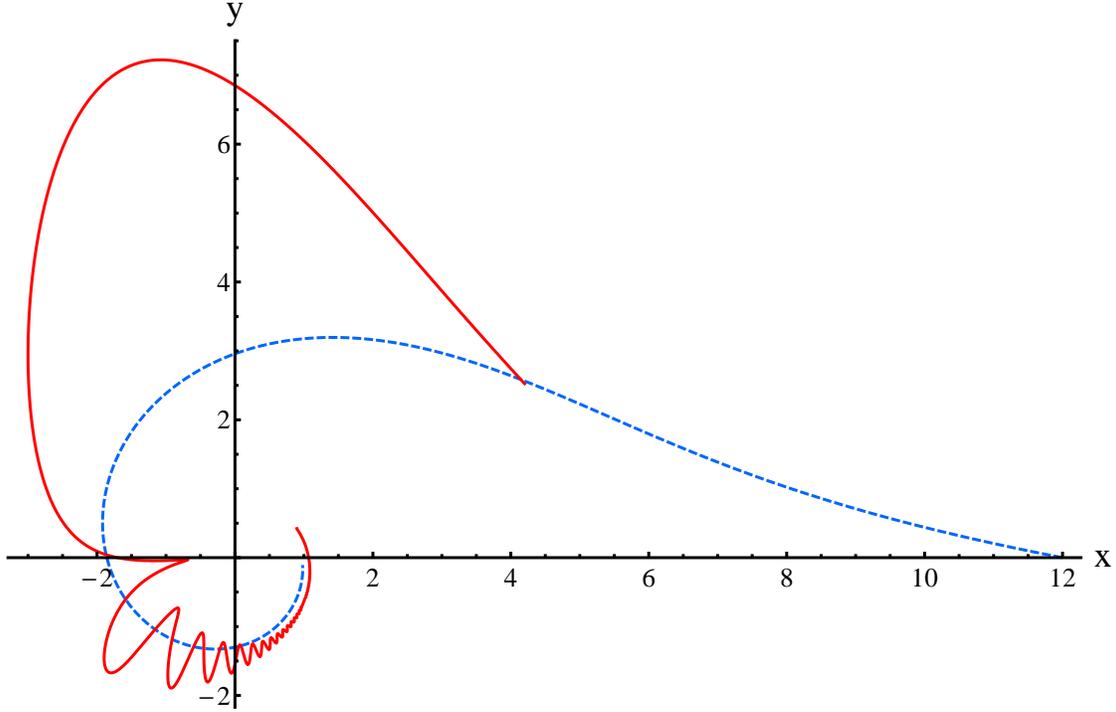}

\caption{ 
Zero approximation trajectory $\vec{r}$ (blue dashed line) and
    Jacobi approximation of the most probable trajectory
    $\vec{r}+\delta\vec{r}$ (red solid line).}
    \end{center}
\end{figure}

The curvature $K$ in the coordinate space for the trajectory $(x(t),y(t))$
    can be defined by the following formula \cite{Kobayashi-Nomizu}:
\begin{equation}\label{GrindEQ__8_} K(t)=\frac{\dot{x}\ddot{y}
    -\dot{y}\ddot{x}}{(\dot{x}^2 +\dot{y}^2)^{3/2} }.\end{equation}
The dependence of the curvature $K(t)$ on the time $t$ is represented in Fig. 3a.
The four insets into this figure illustrate scale-invariant fragments of the dependence
    of $K(t)$ \eqref{GrindEQ__8_}, which reveal its fractal-like behavior.
One then may examine a set  of states with null curvatures $K$ and, respectively, a practically infinite number
of inflexion points of the curve, at     sufficiently large times.
The infinite number of inflexion points implies the coexistence of liquid-expanded and
    crystalline phases.
Hence, the monolayer is in a metastable state during the first-order phase transition.
\begin{figure}[ytbp]
\begin{center}(a)\hspace{7cm}(b)\\

    \includegraphics[width=11cm,height=8cm]{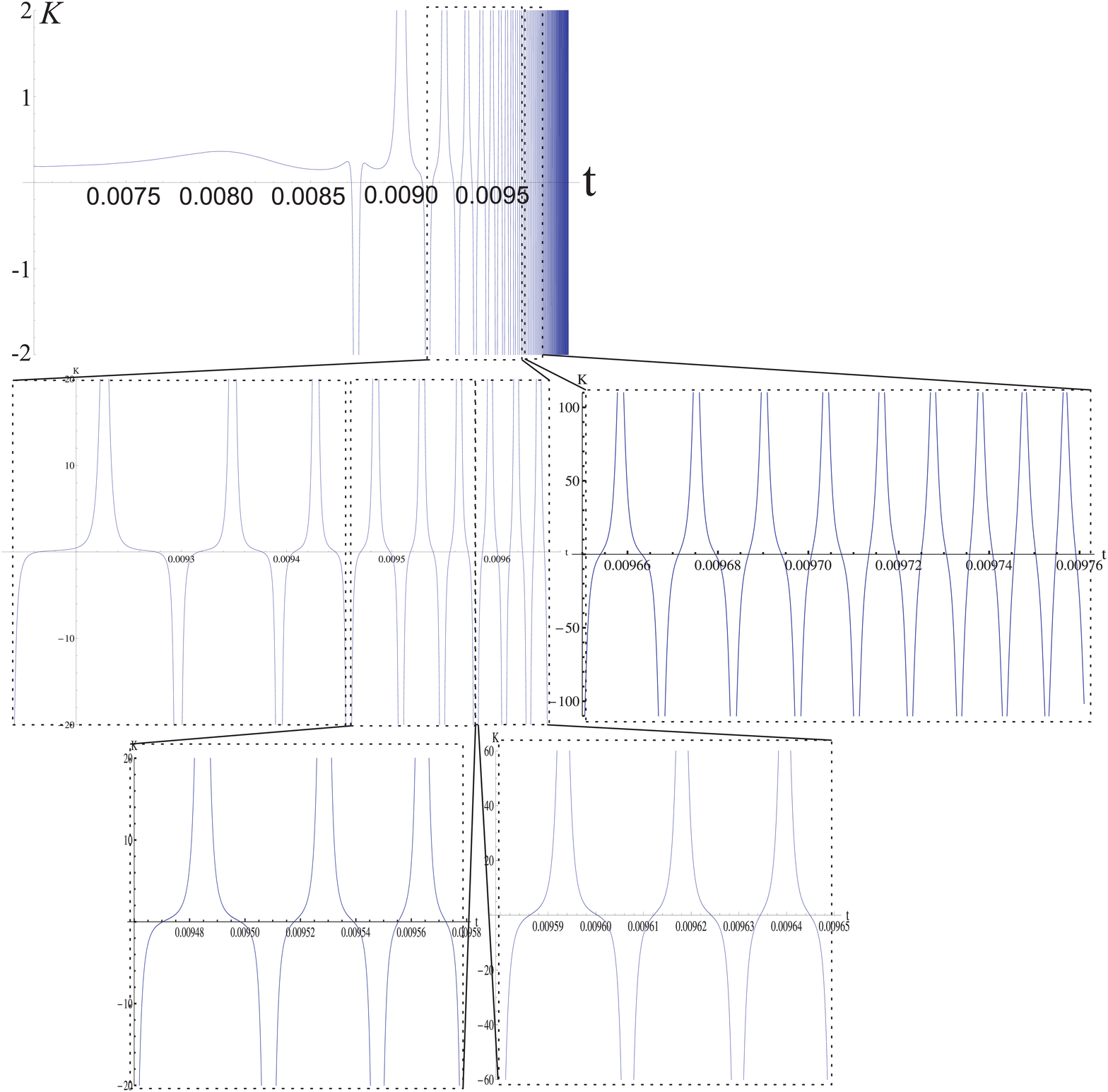}
    \includegraphics[scale=.23]{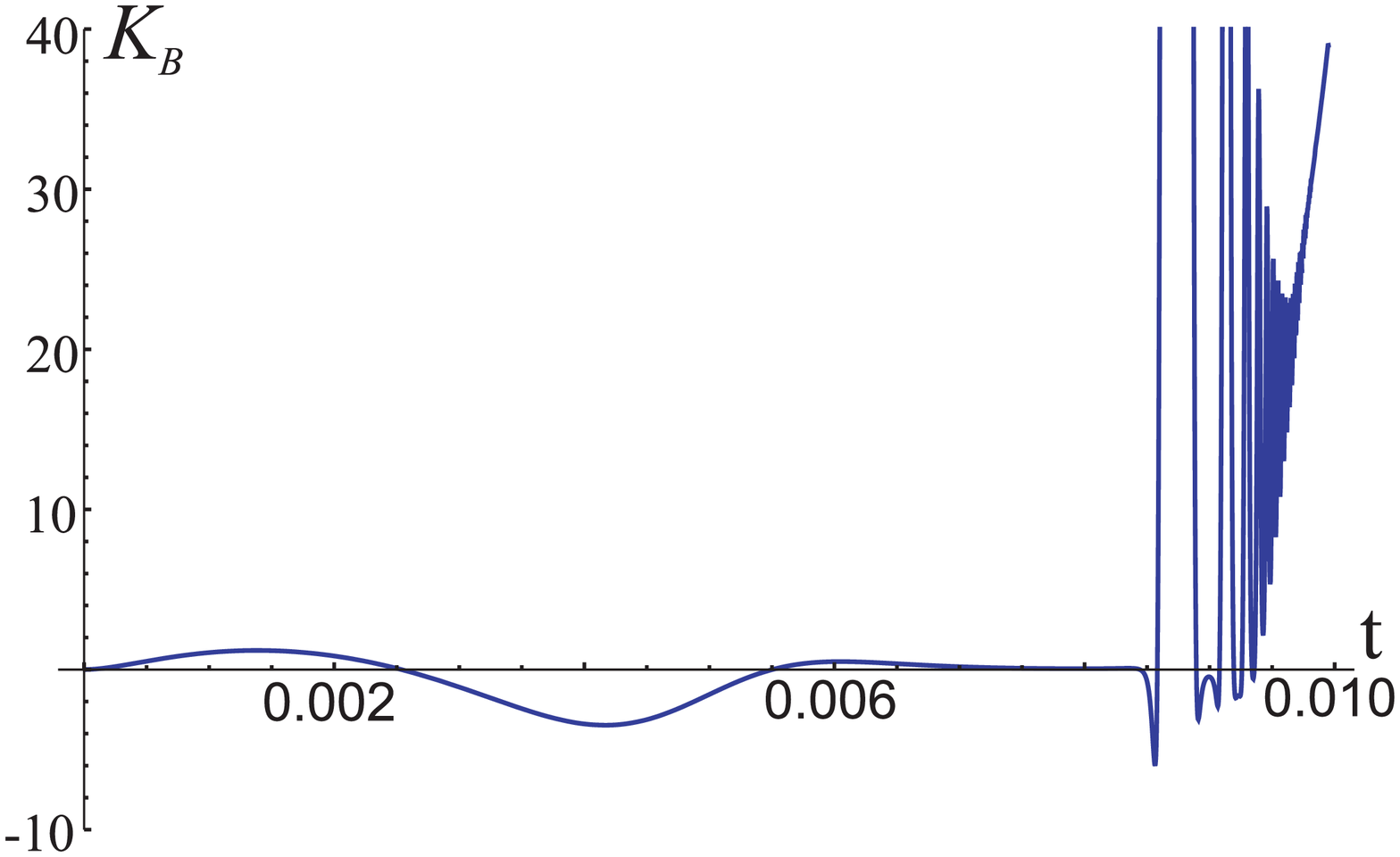}\\
\caption{ 
Time dependence of curvature $K$ (a) and of flag curvature $K_B$
(b).
    Four insets into figure a illustrate a scale-invariant behavior of the curvature $K$.}
    \end{center}
\end{figure}
We further calculate the Lagrange-Berwald curvature $K_B$ on the
tangent bundle,   which is a natural Lagrangian extension $K$ of the {\em flag curvature}
    from Finsler space theory (see \cite[p. 54]{Miro+Hrim+Shim+Saba}).
The flag curvature $K_B$ of the monolayer is determined by the
following formula \cite{17}:
\begin{eqnarray}K_{B}:=K\left( t,r,\varphi ,\dot{r},\dot{\varphi};\ X\right)
    \overset{def}{=}\frac{B_{hijk}y_{1}^{h}X^{i}y_{1}^{j}X^{k}}{\left(g_{hj}g_{ik}
    -g_{hk}g_{ij}\right) y_{1}^{h}X^{i}y_{1}^{j}X^{k}},\label{Berwald-Lagrange-curvature}
\end{eqnarray}
considered with respect to the vector field $X$:
$$X(t,r,\varphi )=X^1
\frac{\partial }{\partial r}+X^2 
    \frac{\partial }{\partial \varphi }\neq 0,$$
where
    $B_{hijk}=g_{hs}\dfrac{\partial^{3} G^{s}}{\partial y_{1}^{i}
    \partial y_{1}^{j} \partial y_{1}^{k}}$. 
The flag curvature $K_B$  along the trajectory $\vec{\tilde
r}=\vec{r} +\delta\vec{r}$
    with the radial flag
$$\{\tilde y_{1},\ X\}=\left\{\tilde y_{1},\left(X^1 
, 0\right)\right\},\quad\tilde y_{1}
    =(\dot {\tilde r}_{0},\dot {\tilde \varphi})=(\dot r
    + \delta \dot r, \dot \varphi+ \delta \dot \varphi)$$
is equal to
\begin{eqnarray}K_{B}=\left[ \frac{18}{\tilde{p} \tilde r^{6}|V|
    e^{\frac{2|V|t}{\tilde r}}}-\frac{4 {U}(t,\tilde r)}{m \tilde r^{2}
    \dot{\tilde \varphi}^{2}}\right] \cdot \dot{\tilde r}^{2}\ \ .\label{flag_curvature}
\end{eqnarray}
The result of the calculation of $K_B$ \eqref{flag_curvature} is represented in Fig.~3b.
By comparing figures 3a and 3b, we conclude that $K_B$ becomes non-positive
    earlier than the curvature $K$ does.
Thus, the loss of stability of the monolayer state begins with the
change of sign
    of the flag curvature $K_B$ and leads to a fractal-like distribution of phase
    elements in the monolayer space.
After the phase transition was accomplished, the fractal time
structure within the
    monolayer is maintained during a certain time interval.
%
%
\section{ Discussion and conclusions}
By using the thermodynamics of non-stationary processes, we
describe an entropy evolution
    of  microstates, and determine a statistical manifold, on which a
    multiple-relaxation-time dynamics occurs.
We propose a continuous parameterization of the first-order phase
transition (a
    {\em heterogeneous dynamics}) for compressed monolayers of amphiphilic molecules
    on which the electrocapillary forces act.
The information geometrodynamic approach with the electrocapillary
forces regarded as
    an information constraint is then applied to describe the phase transition from a
    liquid expanded state into a tilted condensed state with a time fractality for the
    monolayers.
The self-similarity of ordinary coordinate curvature $K $ is an
appearance of the time
    scaling for arbitrary time domain.\par
The flag curvature $K_B$ of the tangent bundle changes its sign
from "$+$"\ to "$-$"\
    already at small compression times, when the apparent $K$ still is positive.
The negative $K_B$ and, respectively, the non-stability of
hydrated complexes in a
    compressed monolayer leads to an escape of water molecules from the complex,
    resulting in the disappearing of steric hindrances to interact amphiphilic
    molecules each other.
Therefore, the phenomenon of precise miscibility of blended
Langmuir monolayer
    \cite{Stereocomplexation}, that is fabricated from apparently immiscible
    substances may proceed from the non-stability.\par
Hence, the Finsler geometry approach allows to analyze the
structural stability loss,
    leading to structuring of transient processes in compressed monolayers.

\section{Acknowledgements }
The present work was developed under the auspices of the Project RA-4.2 /14.06.2014 - BRFFR-RA F14RA-006, within the cooperation
framework between Romanian Academy and Belarusian Republican
Foundation for Fundamental Research.

\end{document}